\documentclass[twocolumn,showpacs,preprintnumbers,amsmath,amssymb]{revtex4}
\usepackage{graphicx}
\usepackage{dcolumn}
\usepackage{bm}

\usepackage{subeqnarray}
\usepackage[dvips]{epsfig}
\usepackage{epic}
\usepackage{natbib}

\begin{document}

\preprint{APS/123-QED}

\title{ Onset of unsteady horizontal convection in rectangle tank at $Pr=1$ }

\author{ Liang SUN }
\email{sunl@ustc.edu.cn;sunl@ustc.edu} \affiliation{School of
Earth and Space Sciences,
 University of Science and Technology of China, Hefei, 230027, China.}
\affiliation{LASG, Institute of Atmospheric Physics, Chinese
Academy of Sciences, Beijing 100029, China.}

\author{Dong-Jun MA}
\affiliation{Dept. of Modern Mechanics,
 University of Science and Technology of China, Hefei, 230027, China.}

\author{ Wei  ZHANG}
\affiliation{Dept. of Modern Mechanics,
 University of Science and Technology of China, Hefei, 230027, China.}

\author{De-Jun SUN }
\affiliation{Dept. of Modern Mechanics,
 University of Science and Technology of China, Hefei, 230027, China.}
\begin{abstract}

The horizontal convection within a rectangle tank is numerically
simulated. The flow is found to be unsteady at high Rayleigh
numbers. There is a Hopf bifurcation of $Ra$ from steady solutions
to periodic solutions, and the critical Rayleigh number $Ra_c$ is
obtained as $Ra_c=5.5377\times 10^8$ for the middle plume forcing
at $Pr=1$, which is much larger than the formerly obtained value.
Besides, the unstable perturbations are always generated from the
central jet, which implies that the onset of instability is due to
velocity shear (shear instability) other than thermally dynamics
(thermal instability). Finally, Paparella and Young's  first
hypotheses about the destabilization of the flow is numerically
proved, i.e. the middle plume forcing can lead to a
destabilization of the flow.
\end{abstract}
 \maketitle

Horizontal convection, in which the water is unevenly heated at
the horizontal surface, was taken as a model of abyssal ocean
circulation. As the abyssal ocean circulation plays an important
role in climate change, the horizontal convection has intensively
been explored in recent years
\cite[]{Paparella2002,Mullarney2004,WangWei2005}. It can be set to
motion by any small temperature gradient, unlike the
Rayleigh-B\'{e}nard convection. But similar to Rayleigh-B\'{e}nard
convection, the horizontal convection may be unsteady at high
Rayleigh numbers $Ra$. There is a critical Rayleigh number $Ra_c$,
and the steady flow is unstable and becomes unsteady when
$Ra>Ra_c$. The unsteady flow in horizontal convection was first
found by numerical simulation \cite[]{Paparella2002}, then was
observed in the experiment at $Ra>10^{12}$ \cite[]{Mullarney2004}.
This unsteady flow is proved to be non-turbulent even as
$Ra\rightarrow \infty$, though the flow field seems to be chaotic
\cite[]{Paparella2002}. The investigation on the unsteady
horizontal convection flow is relatively less, except for
\cite[]{Paparella2002,Mullarney2004,Hughes2007}. However, they
have mainly focused on how the turbulent plume maintains a stable
stratified circulation. Yet how the horizontal convection turned
to be unsteady remains an elusive problem.

To understand this problem, both $Ra_c$ for the onset of unsteady
flow and instability mechanism are of vital. Paparella and Young
\cite{Paparella2002} found $Ra_c\approx 2\times 10^{8}$ at $Pr=1$
in their simulations, which is significantly smaller than others'
results
\cite[]{Rossby1965,Mullarney2004,WangWei2005,Hughes2007,Rossby1998,Siggers2004,SunL2006_jhd1}.
For example, Rossby (1965), Wang and Huang (2005) found the flow
is steady and stable for $Ra<5\times 10^{8}$ in their experiments
\cite[]{Rossby1965,WangWei2005}. Yet other numerical simulations
\cite[]{Rossby1998,Siggers2004,SunL2006_jhd1} have not found
unsteady flows for $Ra<10^9$. Paparella and Young
\cite{Paparella2002} explained the difference of their results
from others' as: (i) middle plume forcing in the numerical
simulations instead of sidewall plume forcing in the experiments,
and (ii) lower aspect ratio ($H/L=1/4$) in their simulations. Both
may lead to destabilization of the flow at lower Rayleigh numbers.
However, their hypotheses have not been intensely investigated.
According to a recent investigation, the flow is still stable for
$Ra<10^{11}$ even at a much lower aspect ratio ($H/L=1/10$)
\cite[]{SunL2007aps}. Thus, it maybe the middle plume forcing that
leads to destabilization at lower Rayleigh numbers.

Our interest here is to verify their first hypotheses. Is the flow
with middle plume forcing less stable than the sidewall plume
forcing? How does the instability occur? To investigate these
problems, more accurate numerical prediction of $Ra_c$ is need for
both forcing cases, for the spatial resolution of simulation was
very lower in the past (e.g. $128\times32$ coarse meshes are used
in \cite{Paparella2002}). Then the flow field under both middle
and sidewall plume forcings are compared, which leads to an
affirmative answer of the above problem.

\begin{figure}
\centerline{
  \includegraphics[width=10cm]{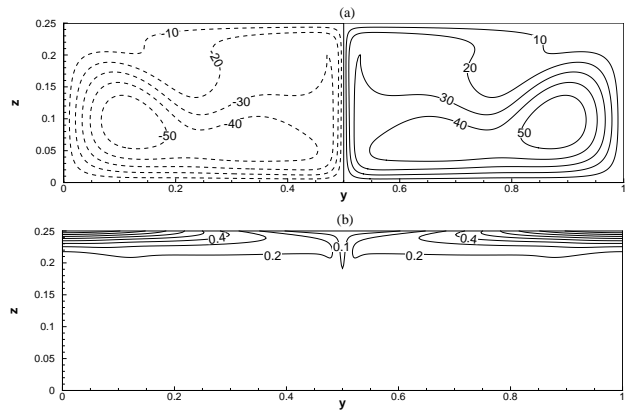}}
  \caption{The flow stream function (a) and temperature field (b) of $Ra=5\times 10^{8}$.
  It is steady and stable and symmetric with middle plume forcing, solid and dashed
  curves for positive and negative values, respectively.}
\label{Fig:HConv_Pr1H025_Ra5E8_PsiDen}
\end{figure}

Similar to the previous investigations, we consider the horizontal
convection flows within the two-dimensional domain, and the
Boussinesq approximation is assumed to be valid for these flows.
As shown in Fig.\ref{Fig:HConv_Pr1H025_Ra5E8_PsiDen}, the
horizontal (y) and vertical (z) regimes are $0\leq y \leq L$ and
$0\leq z\leq H$, respectively. Similar to \cite{Rossby1965}, the
depth $L$ is taken as reference length scale and $A=H/L=1/4$
denotes the aspect ratio. Taking account of divergence-free of
velocity field in Boussinesq approximation, the Lagrangian
streamfunction $\Psi$ and the corresponding vorticity $\omega$ are
introduced. The velocity $\overrightarrow{\mathrm{u}}=(v,w)$,
where horizontal velocity $v=\frac{\partial \Psi}{\partial z}$ and
vertical velocity $w=-\frac{\partial \Psi}{\partial y}$,
respectively. The governing equations in vorticity-streamfunction
formulation are \cite[]{Quon1992,Paparella2002,Siggers2004}:

\begin{subeqnarray} \frac{\partial T}{\partial t} + J(\Psi,T) &=&
(\frac{\partial^2 T }{\partial y^2}+\frac{\partial^2
T }{\partial z^2})\\
\frac{\partial \omega}{\partial t} + J(\Psi,\omega) &=-& Pr
(\nabla^2 \omega+  Ra \frac{\partial
T}{\partial y})\\
 \nabla^2  \Psi&=&-\omega
 \label{Eq:thermo_ctl_Horizontal}
 \end{subeqnarray}
where $J(\Psi,\phi)=\frac{\partial \Psi}{\partial y}\frac{\partial
\phi}{\partial z}-\frac{\partial \phi}{\partial y}\frac{\partial
\Psi}{\partial z}$ denotes the nonlinear advection term. There are
two important dimensionless parameter in
Eq.(\ref{Eq:thermo_ctl_Horizontal}), i.e. Rayleigh number
$Ra=\alpha_T \Delta T gL^3/(\kappa \nu)$ and Prandtl number
$Pr=\nu/\kappa$, where $g$, $\alpha_T$, $\Delta T$, $L$, $\kappa$
and $\nu$ are gravity acceleration, thermal expansion coefficient,
surface temperature difference, length of horizontal domain,
thermal diffusivity and kinematic viscosity, respectively.
Alternatively, Paparella and Youngs used vertical length $H$ as
length scale, so $Ra=64Ra_H$, where $Ra_H$ is the vertical
Rayleigh number by using vertical length $H$ as unit
\cite[]{Paparella2002}.

\begin{figure}
\centerline{
  \includegraphics[width=10cm]{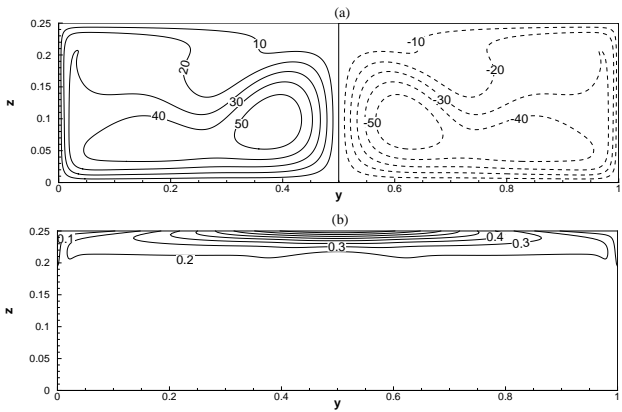}}
  \caption{The flow field (a) and temperature field (b) of $Ra=5\times
  10^{8}$,
  which are steady and stable and symmetric with sidewall plume forcing.}
\label{Fig:HConv_Pr1H025_Ra5E8_PsiDen2}
\end{figure}

More specifically, we consider the horizontal convection in a
rectangle tank at $Pr=1$. The tank has same velocity boundary
condition as that in \cite{Paparella2002}, i.e. free slip and no
shear stress at the walls. In addition, two different surface
forcings are used, which are central symmetric. One is middle
plume forcing as $T=[1+\cos(2\pi y)]/2$ \cite{Paparella2002}, the
other is sidewall plume forcing as $T=[1-\cos(2\pi y)]/2$
\cite{Quon1992,SunL2007aps}. Comparing these with one cell forcing
$T=\cos(\pi y/2)$ \cite{Rossby1998}, there are two symmetric cells
in the flow field under such forcings (e.g.
Fig.\ref{Fig:HConv_Pr1H025_Ra5E8_PsiDen} and
Fig.\ref{Fig:HConv_Pr1H025_Ra5E8_PsiDen2}), when the flow is
symmetrically steady and stable. In addition, the middle plume
forcing in the left cell is the same with the sidewall plume
forcing in the right cell (see
Fig.\ref{Fig:HConv_Pr1H025_Ra5E8_PsiDen} behind). Thus in the
steady flows, both forcings will lead to the same flow patterns
except for a position shift, which is proved by the following
investigation.

There are two important quantity describing the circulation, i.e.
the non-dimensional streamfunction maximum and the non-dimensional
heat flux. The non-dimensional streamfunction maximum
$\Psi_{\max}=\Psi^*_{\max}/\nu$, where $\Psi^*_{\max}$ is the
maximum of the dimensional streamfunction.

\begin{table}
 \caption{\label{Table:NatConv_Benchmark} Comparison of
the benchmark solutions from  \cite{Lequere1991} and
\cite{TianZF2003}, $Pr=0.71$. $\Psi_{mid}$, $\Psi_{\max}$ are the
values in the midpoint and the maximum of streamfunction,
respectively. And $Nu$ is average Nusselt number at the heated
wall. The resolution is $80\times80$ meshes for present results.}
\begin{ruledtabular}
\begin{tabular}{ccccc}
author & $Ra$ & $\Psi_{mid}$ & $\Psi_{\max}$ &  $Nu$ \\
\hline
\cite{Lequere1991} & $10^6$ & $16.386$ & $16.811$  &$8.822$ \\
\cite{TianZF2003} & $10^6$ & $16.386$ & $16.811$ &$8.825$ \\
Present & $10^6$ & $16.430$ & $16.863$  &$8.828$ \\
\cite{Lequere1991} & $10^7$ & $29.361$ & $30.165$  &$16.523$ \\
\cite{TianZF2003} & $10^7$ & $29.356$ & $30.155$  &$16.511$ \\
Present & $10^7$ & $29.605$ & $30.448$  &$16.535$ \\
\end{tabular}
\end{ruledtabular}
\end{table}

The above Eq.(\ref{Eq:thermo_ctl_Horizontal}) is solved with
finite different method in non-uniform grids. Crank-Nicholson
scheme and Arakawa scheme \cite[e.g.][]{Arakawa1966,Olandi2000}
are applied to discretize the linear and nonlinear terms,
respectively. Comparing to the other schemes, Arakawa scheme is
more accurate but more expensive, and it has also been applied to
horizontal convection flows at high Rayleigh number
\cite[]{SunL2006_jhd1,SunL2007aps}. Table
\ref{Table:NatConv_Benchmark} shows the validation of the scheme
with nature convection problem. A fine spatial resolution mesh of
$512\times128$ is used to eliminate numerical instability.

\begin{figure}
\centerline{
  \includegraphics[width=10cm]{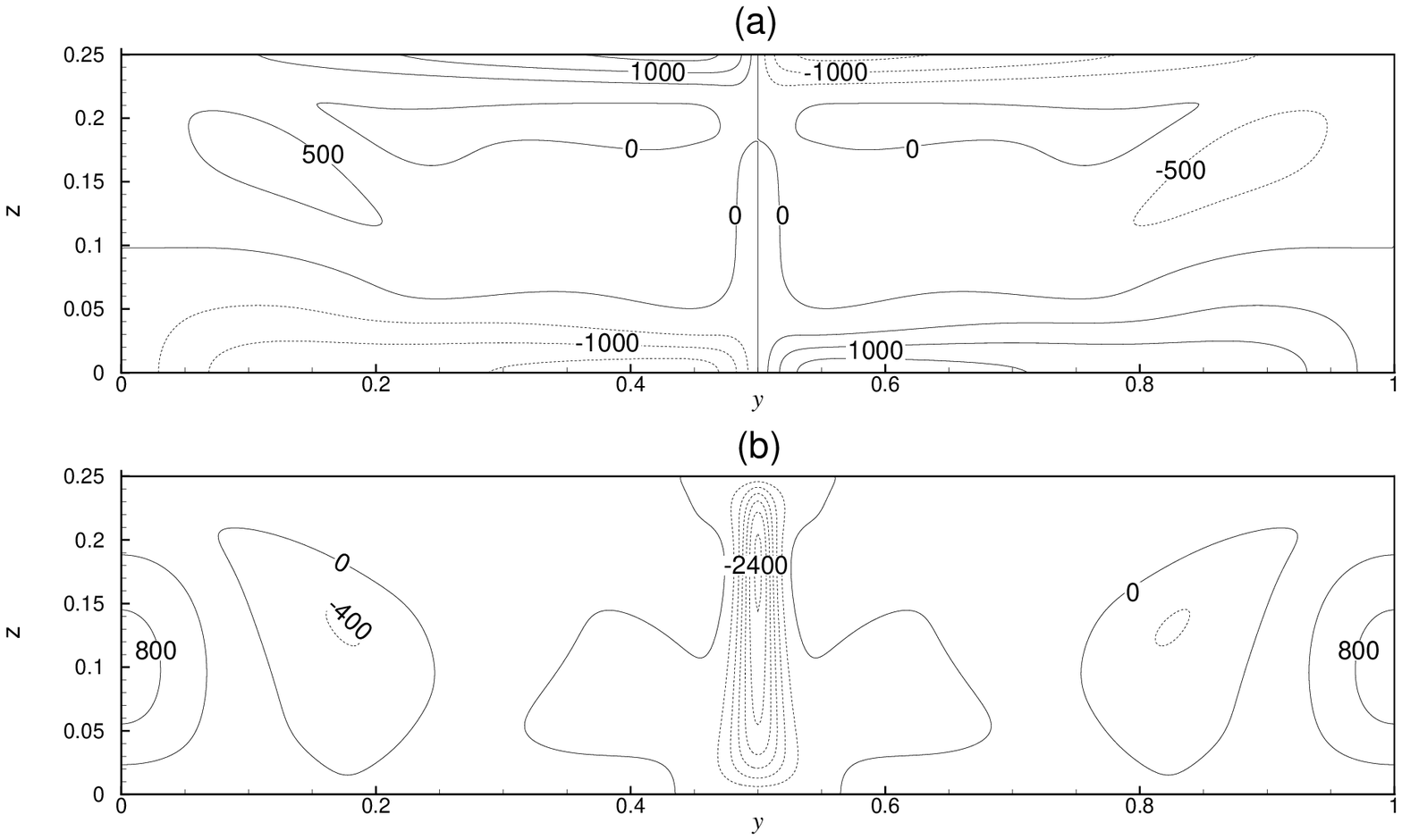}}
  \caption{The horizontal (a) and vertical (b) velocity fields of $Ra=5\times 10^{8}$.
  It is steady and stable and symmetric with middle plume forcing, solid and dashed
  curves for positive and negative values, respectively.}
\label{Fig:HConv_Pr1H025_Ra5E8_VW}
\end{figure}

First, the middle plume forcing is considered, which is steady and
stable for $Ra<5.5\times 10^8$.
Fig.\ref{Fig:HConv_Pr1H025_Ra5E8_PsiDen} shows the flow field (a)
and temperature field (b) of $Ra=5\times 10^{8}$ with
$\Psi_{\max}=59.83$, in which the flow is symmetric, steady and
stable. In this case, the center line symmetrically separates the
flow field into two parts, like a free slip wall. There is a
vigorous downward jet in the center of tank corresponding to the
middle plume forcing (Fig.\ref{Fig:HConv_Pr1H025_Ra5E8_PsiDen}a),
where the vertical velocity field has a minimum of $w=-2513$
(Fig.\ref{Fig:HConv_Pr1H025_Ra5E8_VW}b). The center jet leads to
the clockwise and anticlockwise plume cells in the left and right
part of tank, respectively. In the left circulation cell, the flow
sinks quickly along the center line and upwells clockwise along
the left side wall with relatively slower speed, which can be also
seen from the vertical velocity of the flow
(Fig.\ref{Fig:HConv_Pr1H025_Ra5E8_VW}). Besides, there are two
horizontal jets respectively near top and bottom walls in the left
circulation cell (Fig.\ref{Fig:HConv_Pr1H025_Ra5E8_VW}a). Totally,
there are 2 horizontal jets near wall and a vertical jet at the
center in each cell. Contract to the flow field, the temperature
field is very simple. An obvious boundary layer exists near the
surface in temperature field, which leads to a 1/5-power law of
$Ra$ for heat flux \cite[e.g.][]{Rossby1965,Quon1992,Siggers2004}.
And below the temperature boundary layer, the temperature is
almost homogeneous due to the convection. Thus there is a very
strong stratification near the surface ($\partial T/\partial z\sim
Ra^{1/5}$) but a very weak stratification in other region
($\partial T/\partial z\sim 0$). As the above case is stable, so
the critical Rayleigh number must be larger than $5\times10^{8}$,
which is significantly larger than the value obtained before
\cite{Paparella2002}.

\begin{figure}
\centerline{
  \includegraphics[width=6cm]{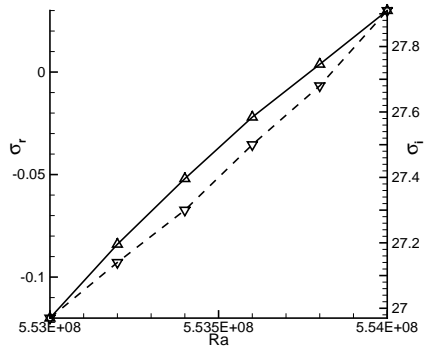}}
  \caption{Growth rate $\sigma_r$ (solid) and $\sigma_i$ (dashed) vs. $Ra$, respectively.
  \label{Fig:HConv_Pr1H025_CrCi_Ra}}
\end{figure}

To find the critical Rayleigh number $Ra_c$, the growth rate of
perturbation $\phi(t)$ is calculated numerically. And $\phi(t)$ is
assumed to satisfy $\phi(t)=e^{\sigma t}\phi(0)$, where
$\sigma=\sigma_r+i\sigma_i$ is the complex growth rate of
disturbance. It is found that the onset of unsteady flow is at
$Ra_c=5.5377\times 10^8$, as shown in
Fig.\ref{Fig:HConv_Pr1H025_CrCi_Ra}. For $Ra=5.53\times 10^8$, the
flow is stable and the growth rate is approximately
$\sigma_r=-0.12$. But the flow is unstable and the growth rate is
approximately $\sigma_r=0.03$ for $Ra=5.54\times 10^8$. Thus the
critical Rayleigh number $Ra_c$ is obtained $5.53\times
10^8<Ra_c<5.54\times 10^8$. The accurate value of
$Ra_c=5.377\times 10^8$ is obtained by interpolating from the
above result. Moreover, the onset of unsteady flow is found to
occur via Hopf bifurcation. As Fig.\ref{Fig:HConv_Pr1H025_CrCi_Ra}
shows, the image part of growth rate is nonzero and the eigenmode
of perturbation is periodic. This Hopf bifurcation of the
horizontal convection has not been reported yet, and previous
investigations dealt only with chaotic flows.

\begin{figure}
\centerline{
  \includegraphics[width=10cm]{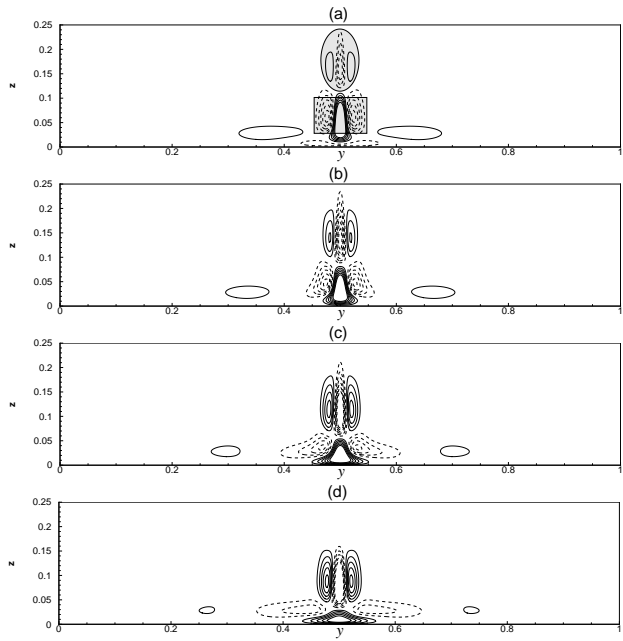}}
  \caption{The perturbational vorticity fields at $t=0$ (a), $t=T/8$ (b),
$t=T/4$ (c) and $t=3T/8$ of $Ra=5.54\times 10^{8}$, solid and
dashed curves for positive and negative values, respectively.}
  \label{Fig:HConv_Pr1H025_Ra540E8_V03}
\end{figure}

Meanwhile, the evolution of the perturbation vorticity fields
during the first half period at $t=0$ (a), $t=T/8$ (b), $t=T/4$
(c) and $t=3T/8$ of $Ra=5.54\times 10^{8}$ are depicted in
Fig.\ref{Fig:HConv_Pr1H025_Ra540E8_V03}, respectively. The
perturbation vorticity fields are symmetric about centerline,
which implies that the horizontal velocity is nonzero at
centerline. It can be seen that the perturbation tripole A (the
shadowed ellipse in Fig.\ref{Fig:HConv_Pr1H025_Ra540E8_V03}a) is
generated from central downward jet, then propagates and amplifies
along the central jet downward to the bottom wall
(Fig.\ref{Fig:HConv_Pr1H025_Ra540E8_V03}b,c,d). When tripole A
approaching to the bottom, it becomes weaker and weaker and breaks
into two parts: the left and the right near the bottom, which can
be seen from the evolution of tripole B (the shadowed rectangle in
Fig.\ref{Fig:HConv_Pr1H025_Ra540E8_V03}a). And the mean flow
advects the broken vortexs horizontally along the bottom wall
(Fig.\ref{Fig:HConv_Pr1H025_Ra540E8_V03}b,c,d). Then in the second
half period, a reverse tripole will generate  the same place of
vortexes A at $t=T/2$, and the same story repeats for it, which is
omitted here. In short, the perturbations generate and amplify in
the central vertical jet, but are propagated and weakened along
the horizontal wall.

\begin{figure}
\centerline{
  \includegraphics[width=10cm]{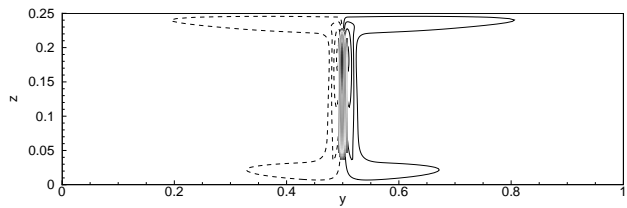}}
  \caption{The vorticity of $Ra=5.54\times 10^{8}$ with vertical velocity $w$ (shadowed as $w>1800$),
solid and dashed curves for positive and negative values,
respectively.}
  \label{Fig:HConv_Pr1H025_Ra540E8_VorW}
\end{figure}

Further investigation shows that the instability of flow occurs
due to shear. First, as we noting, the instabilities always occur
in the center and propagate along the mean flow. Second, this
trigger place locates in the area where there is a vigorous jet
with strong shear (see e.g. Fig.\ref{Fig:HConv_Pr1H025_Ra5E8_VW}b
and Fig.\ref{Fig:HConv_Pr1H025_Ra540E8_VorW})
 (Fig.\ref{Fig:HConv_Pr1H025_Ra540E8_VorW}).  As the
stratification is very weak here (see e.g.
Fig.\ref{Fig:HConv_Pr1H025_Ra5E8_PsiDen}b), the flow in this
region is dominated by momentum dynamics other than thermal
dynamics. All these imply that the onset of instability leading to
unsteady flow is due to shear instability at larger Rayleigh
numbers, which is much different from Rayleigh-B\'{e}nard
instability. However, shear is not the sufficient condition for
instability. For example, the instability near the top surface is
suppressed due to strong stratification
(Fig.\ref{Fig:HConv_Pr1H025_Ra5E8_PsiDen}b), though both the
velocity (Fig.\ref{Fig:HConv_Pr1H025_Ra5E8_VW}a) and the shear
(Fig.\ref{Fig:HConv_Pr1H025_Ra540E8_VorW}) near top surface are
still very large. In words, the onset of instability is due to
velocity shear (shear instability) other than thermally dynamics
(thermal instability).

Now, the sidewall plume forcing is considered.
Fig.\ref{Fig:HConv_Pr1H025_Ra5E8_PsiDen2} shows the flow field and
temperature field of $Ra=5\times 10^{8}$, in which the flow is
symmetric, steady and stable like that under middle plume forcing.
There are two vigorous downward jets near the walls corresponding
to the sidewall plume forcing
(Fig.\ref{Fig:HConv_Pr1H025_Ra5E8_PsiDen2}a). As mentioned above,
the sidewall plume forcing will lead to exactly the same flow
pattern as the middle plume forcing does except for a position
shift, which can be seen from
Fig.\ref{Fig:HConv_Pr1H025_Ra5E8_PsiDen} and
Fig.\ref{Fig:HConv_Pr1H025_Ra5E8_PsiDen2}. As the flow is stable,
the center line like a free slip wall symmetrically separates the
two cells. The left cell in
Fig.\ref{Fig:HConv_Pr1H025_Ra5E8_PsiDen} is exactly the same with
the right cell in Fig.\ref{Fig:HConv_Pr1H025_Ra5E8_PsiDen2}.
However, the flow is much more stable with the sidewall plume
forcing. And the critical Rayleigh number is found $Ra_c\approx
1.85\times 10^{10}$ (with  $768\times 192$ meshes) in this case.
As mentioned above, the flow is much more stable with the sidewall
plume forcing than that with the middle plume forcing, though both
forcings lead to the same flow patterns. This is very interesting,
and can be understood from the mechanism of instability.

It's found that the rigid wall suppresses the perturbation, which
leads a more stable flow with the sidewall plume forcing than that
with the middle plume forcing. As the loss of stability is due to
strong velocity shear in the center in horizontal convection, the
smaller the shear is, the more stable the flow is. In the case of
middle plume forcing, the perturbation with nonzero horizontal
velocity occurs at the most vigorous downward jet. And the
perturbed flows cross the center line and propagate downstream.
However, in the case of sidewall plume forcing, these crossing
flows are suppressed by rigid walls. So that the critical Rayleigh
number is much larger in this case. Paparella and Young (2002)
hypothesized that middle plume forcing may lead to a
destabilization of the flow. Here this hypotheses is proved both
physically and numerically.

In conclusion, the onset of unsteady flow is found to occur via a
Hopf bifurcation in the regime of $Ra>Ra_c=5.5377\times 10^8$ for
the middle plume forcing at $Pr=1$, which is much larger than the
previously obtained value. Besides, the onset of unsteady flow is
due to shear instability of central downward jet. Finally, the
first hypotheses of Paparella and Young (2002) for instability is
numerically approved, i.e. the middle plume forcing can lead to a
destabilization of the flow at relatively lower Rayleigh numbers.

This work is supported by the National Basic Research Program of
China (No. 2007CB816004), the National Foundation of Natural
Science (No. 40705027, No. 10602056 and No. 10772172), the
National Science Foundation for Post-doctoral Scientists of China
(No. 20070410213), and the Presidential Foundation of the Chinese
Academy of Sciences, China.

\fontsize{10pt}{10pt}\selectfont
\addcontentsline{toc}{chapter}{参考文献}
\addtolength{\itemsep}{-4em} 

\end{document}